\documentclass[doublespacing]{elsart}
\usepackage{graphicx}
\usepackage{graphics}
\usepackage{epsfig}
\usepackage{amssymb}
\begin{document}
\begin{frontmatter}
\title{$2\times250$ GeV CLIC $\gamma\gamma$ Collider Based on it's Drive Beam FEL}
\author[label1,label2]{H\"usn\"u Aksakal\corauthref{label1,label2}}
\corauth[label1,label2]{Corresponding author}
\ead{aksakal@science.ankara.edu.tr}
\address[label1]{Department of Physics, Faculty of Science, Ankara University, 06100 Tandogan, Ankara, Turkey}
\address[label2]{Department of Physics, Faculty of Art and Science, Nigde University, 51200 Nigde, Turkey}
\begin{abstract}
CLIC is a linear $e^+e^-$ ($\gamma\gamma$) collider project which
uses a drive beam to accelerate the main beam. The drive beam
provides RF power for each corresponding unit of the main linac
through energy extracting RF structures. CLIC has a wide range of
center-of-mass energy options from 150 GeV to 3 TeV. The present
paper contains optimization of Free Electron Laser (FEL) using one
bunch of CLIC drive beam in order to provide polarized light
amplification using appropriate wiggler and luminosity spectrum of
$\gamma\gamma$ collider for $E_{cm}$=0.5 TeV. Then amplified laser
can be converted to a polarized high-energy $\gamma$ beam at the
Conversion point (CP-prior to electron positron interaction point)
in the process of Compton backscattering. At the CP a powerful laser
pulse (FEL) focused to main linac electrons (positrons). Here this
scheme described and it is show that CLIC drive beam parameters
satisfy the requirement of FEL additionally essential undulator
parameters has been defined. Achievable $\gamma\gamma$ luminosity is
above $10^{34}$.
\end{abstract}
\end{frontmatter}
% main text
\section{Introduction}
The idea of a $\gamma\gamma$ collider was proposed in the early
1980~\cite{telnov83}. It is well known that due to severe
synchrotron radiation in storage rings, $e^{+}e^{-}$ colliders in
the TeV energy region will be linear. Unlike the situation in
storage rings, in linear colliders each bunch is used only once.
This makes possible the use of electrons for production of
high-energy photons to obtain colliding $\gamma\gamma$ and
$\gamma{e}$ beams. The high energy gamma beam is produced by Compton
backscattering of laser light off the electron beam. At about the
same time, it was suggested that a free-electron laser could be used
as the photon source of the $\gamma\gamma$
collider~\cite{kondratenko}. In 1994, the use of single drive-beam
bunches in a free-electron laser for a $\gamma\gamma$ collider based
on an erlier version of CLIC, which accelerated single main bunches
per pulse, was proposed by R.~Corsini and
A.~Mikhailichenko~\cite{corsini94}. The present paper describes to
use a MOPA (Master Oscillator Power Amplifier) FEL instead of a
conventional laser for Compton backscattering. MOPA FEL is studied
for the case in which of radiation from a master oscillator is
amplified in the FEL amplifier with tapered wiggler. But here we
offer to use a solid state laser as a master laser instead of master
oscillator radiation. FEL has many advantages compare to a
conventional laser i.e: tunability, minimum divergence
etc~\cite{corsini94,saldin95a}. The FEL produces the radiation to
Terawatt level which is the required power for a laser with
wavelength of 1 $\mu$m at a photon linear collider. Several physics
opportunities for $\gamma{e}$ and $\gamma\gamma$ collisions  at the
CLIC are described below. Some examples are also described
in~\cite{NLC,CLICtdr}.
\begin{itemize}
\item A $\gamma\gamma$ collider offers  a unique opportunity for measuring
the two-photon decay width of the Higgs boson, providing a glimpse
of the mass scale beyond the TeV range. Even in the case when the
Higgs boson will be found at $e^{+}e^{-}$ linear colliders, its
properties may be studied in detail only with CLIC photon collider.
\item A $\gamma\gamma$ collider is well suited for searching new
charged particles, such as SUSY particles, leptoquarks, excited
state of electrons, etc. because photons generally couple more
effectively to these particles than do electrons or positrons.
\item A $\gamma\gamma$ or $\gamma{e}$ collider serving as a W-factory,
producing $10^{6}-10^{7}$ Ws/year, allows for a precision study of
gauge boson interactions and a search for their possible anomalies.
\item At $\gamma{e}$ collider charged supersymetric particles
with masses higher than the beam energy could be produced as well as
the structure of photon could be measured.
\end{itemize}
A proposed scheme of CLIC $\gamma\gamma$ collider based on it's
drive beam FEL is shown in Fig~\ref{fig:clicnma}.
\section{Kinematic Background}
Conversion of FEL to high energy $\gamma$ beam at the conversion
point can be scaled with dimensionless $x$
parameter~\cite{telnov83}.
\begin{equation}\label{xparam}
x=4E_{b}\omega_{0}/m^2
\end{equation}
where $m$, $E_{b}$ are electron rest mass and beam energy
respectively, $\omega_{0}$ laser photon energy. The maximum energy
of the backscattered photons $\omega_{max}=x E_{b}/(x+1)$ depends on
the parameter $x$ but the backscattered photons can be lost for
$x>>4.8$ due to $e^{+}e^{-}$ pair creation at the collisions of
produced photons with un-scattered FEL photons (Breit-Wheeler
process). Thus, the optimum value is $x=4.8$, giving the maximum
photon energy $\omega_{max}=0.81 E_b$. Neglecting multiple
scattering, and assuming that the laser profile seen by each
electron is the same, the conversion probability of generating high
energy gamma photons per individual electron can be written
as~\cite{NLC}
\begin{equation}\label{conversion}
p=1-e^{-q}
\end{equation}
If the laser intensity along the axis is uniform the exponent q is
\begin{equation}
q=\frac{A}{A_0}=\frac{\sigma_{c}A}{\omega_{0}\Sigma_{L}}=\frac{\sigma_{c}I\tau_{L}}{\omega_{0}}=\frac{\sigma_{c}P\tau_{L}}{\omega_{0}\Sigma_{L}}
\end{equation}
where A/$\omega_0$ denotes total number of laser photons, $\sigma_c$
is the total Compton cross section is equal to 1.75 $10^{-25}cm^{2}$
for $x=4.8$, $I$ is the laser beam intensity and $\tau_{L}$
($\sqrt{2\pi}\sigma_{L,z}(rms)/c$) is the laser pulse duration,
$\Sigma_{L}=\frac{1}{2}\lambda Z_{R}$ the laser beam cross section
at the focal point and $A$ is the laser pulse energy
($A=I\tau_{L}\Sigma_{L}$). The optimum conversion efficiency
corresponds to q=1 which is reached for a laser pulse energy of
$A=A_{0}=\omega_{0}\lambda Z_{R}/2\sigma_{c}$. In this case one has
p=0.65. Required laser-beam parameters are listed in
Table~\ref{dbparam}. It should be kept in mind that last laser spot
size must be bigger than electron beam transverse size, therefore
last laser beam spot size is defined by final optical system of
laser system (before CP). After the conversion, energy spectrum of
high-energy photons are given as~\cite{Borden92,telnov90}:
\begin{eqnarray}
\frac{1}{\sigma_{c}} \frac{d\sigma_{c}}{d\omega}&=& f(\omega) \nonumber \\
&=&\frac{1}{\sigma_{c}E_{b}}\frac{2\pi\alpha^{2}}{xm_{e}^{2}}[\frac{1}{1-y}+1-y-4r(1-r)-\lambda_{e}\lambda_{\gamma}rx(2r-1)(2-y)]\nonumber\\
\end{eqnarray}
where $y=\omega/E_{b}$, $r=y/[x(1-y)]$, $\lambda_{e}$,
$\lambda_{\gamma}$ electron and laser beam helicities respectively
and $\sigma_{c}$ is the Compton cross section of the laser and the
e-beam, given as:
\begin{eqnarray}
\sigma_{c}=\sigma_{c}^{0}+\lambda_{e}\lambda_{\gamma}\sigma_{c}^{1} \\
\sigma_{c}^{0}=\frac{\pi\alpha^{2}}{xm_{e}^{2}}[(2-\frac{8}{x}-\frac{16}{x^{2}})ln(x+1)+1+\frac{16}{x}-\frac{1}{(x+1)^{2}}] \nonumber \\
\sigma_{c}^{1}=\frac{\pi\alpha^{2}}{xm_{e}^{2}}[(2+\frac{4}{x})ln(x+1)-5+\frac{2}{x}-\frac{1}{(x+1)^{2}}]\nonumber
\end{eqnarray}
By varying the polarization of electron and FEL, the polarization of
the high-energy gamma beam can be tailored to fit the needs of the
gamma-gamma collision experiments. Controlling the polarization is
also important for sharpening the spectral peak in the
$\gamma\gamma$ luminosity. Due to dependence of Compton scattering,
the peak in the luminosity spectrum is significantly enhanced by
choosing the helicity of laser photons to be of the opposite sign to
helicity of the electrons~\cite{telnov83,NLC,telnov90,Borden92}.
Required laser parameters for $p$=0.65 can be define using
equation~\ref{conversion}.  The FEL output radiation is totally
polarized: circularly or linearly for the case of helical or planar
undulator, respectively~\cite{saldin95a,saldin95b,saldin2000}. To
reduce the cost of the laser system only free electron laser can be
used~\cite{saldin93,saldin95a,saldin95b}.
\section{Luminosity Calculation}
\begin{figure}
\rotatebox{0}{\includegraphics[width=0.99\textwidth]{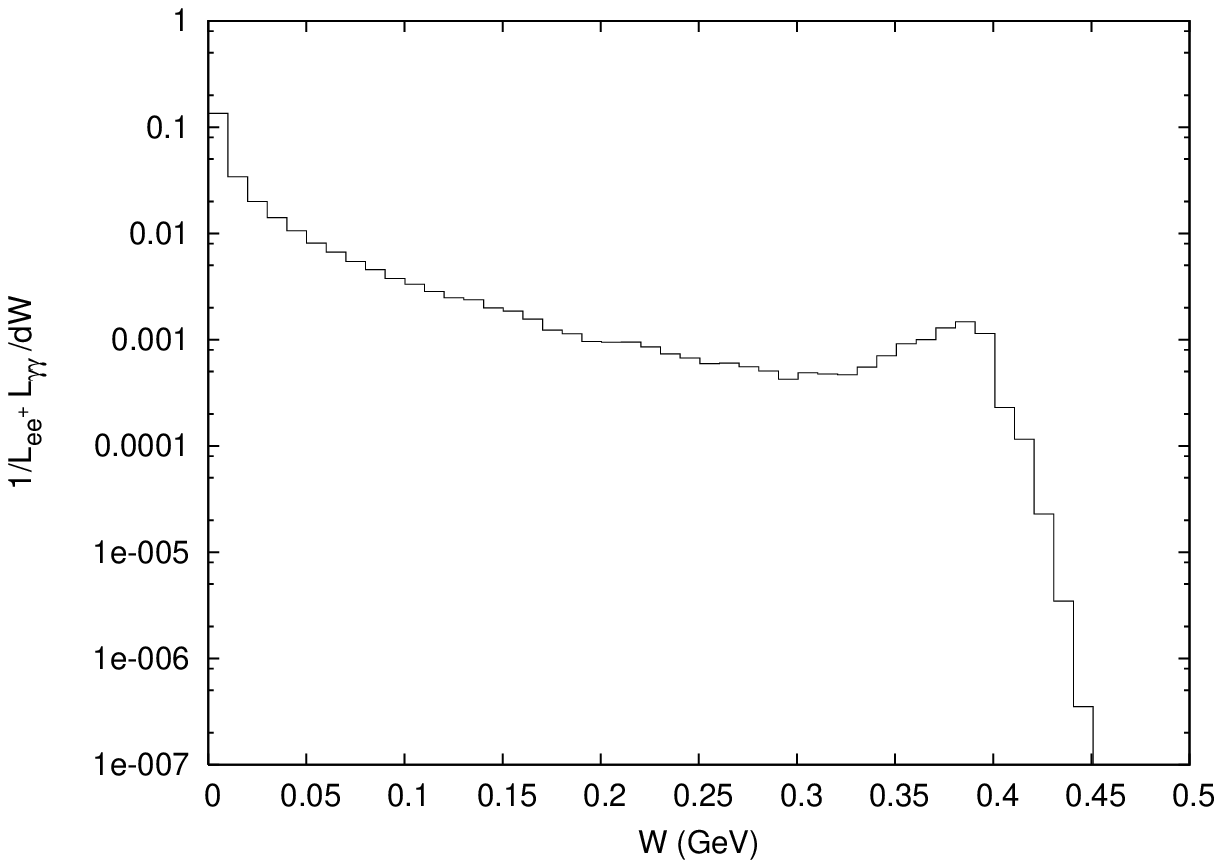}}
\caption{CLIC $\gamma\gamma$ Luminosity spectrum at $E_{cm}$=0.5 TeV
for $L_{ee^{+}}$=5.2 $10^{34}$ $cm^{-2}s^{-1}$} \label{fig:lumi.eps}
\end{figure}

The $\gamma\gamma$ luminosity is approximately proportional to
$e^{+}e^{-}$ geometric luminosity. Luminosity of colliding beams is
related with beam size, number of particle per bunch and repetition
frequency. The transverse beam size of both laser and electron
(positron) beams can be expressed as~\cite{aksakalnimA}:
\begin{equation}
\sigma_{i,j}(s)=\sigma_{i,j}^{\ast}\sqrt{1+\frac{(s-s_{j})^{2}}{{\beta^{\ast}}^{2}}}
\end{equation}
where $j$ represent the beam kind ($e^{-}$, $e^{+}$, $l$) and
$i$($x$,$y$) the transverse coordinate, $\beta^{\ast}$ is the
betatron function at the waist and $s_{j}$ the waist position. The
beam size at the waist is
$\sigma_{i,j}^{\ast}$=$\sqrt{\beta\epsilon_{n}/\gamma}$ where
$\epsilon_{n}$ is normalized beam emittance and $\beta$ is betatron
function. For laser beam the beta function is equal to Rayleigh
range ($Z_{R}$) and the diffraction limited emittance is
$\lambda/4\pi$. After laser optical system the Rayleigh range is
$Z_{R}=\frac{4}{\pi}\lambda F_{N}^{2}$ where $F_{N}$ is defined
roughly as the ratio of focal length to the diameter of focusing
mirror~\cite{NLC}. For high reflective mirrors, the average power
density damage threshold is 10 $MW/cm^2$ and peak power density
damage threshold is 10 $GW/cm^2$ ~\cite{urakawa06}. The problem in
both cases is discharge (breakdown of hydrocarbons on the mirror
surface) on mirror surface.  Average power of required FEL is 33 kW.
Therefore FEL easily transport to CP with mirrors. In our case,
desired value of Rayleigh length (0.263 mm) can be obtained by
taking $F_{N}$ equal to 14 for change the FEL Rayleigh range (1.7
cm) at 1 J pulse energy. For calculation of spectral luminosity CAIN
2.35 simulation program used~\cite{cain}. Nonlinear effect at the
conversion point scaled with dimensionless $\xi^{2}$ parameter,
which is related with photon density in the laser pulse, wavelength,
pulse length, pulse energy. This $\xi^{2}$ parameter is of the
form~\cite{brinkmann97}
\begin{equation}\label{}
\xi^{2}=\frac{4 r_{e}\lambda A}{(2\pi)^{3/2}\sigma_{L,z}mc^{2}Z_{R}}
\end{equation}
In the case of $\xi^{2}$$>>$1 multiphoton process occur at
conversion point, for $\xi^{2}$$<<$1 Compton scattering occur. In
our case $\xi^{2}$=0.3 and there is no problem with multiphoton
process. The differential luminosity equation as a function of laser
energy in a $\gamma\gamma$ collision is given below~\cite{Borden92}:
\begin{eqnarray}
\frac{1}{L_{ee}}\frac{dL_{\gamma\gamma}}{dWd\eta} \nonumber\\
&=&
\frac{W}{2}f_{1}(\frac{We^{\eta}}{2})f_{2}(\frac{We^{-\eta}}{2})I_{0}(\frac{d_{1}d_{2}}{\sigma_{1}(s,s_{e})^{2}+\sigma_{2}(s,s_{e})^{2}})e^{-\frac{d_{1}^{2}+d_{2}^{2}}{2(\sigma_{1}(s,s_{e})^{2}+\sigma_{2}(s,s_{e})^{2})}}\nonumber\\
\end{eqnarray}
where $W=2\sqrt{\omega_{1}\omega_{2}}$ is invariant mass, $I_{0}$
Bessel function of the order of zero,
$\eta=Arctanh(\frac{\omega_{1}-\omega_{2}}{\omega_{1}+\omega_{2}})$
is $\gamma\gamma$ rapidity, $d_{1}=z_{1}\theta_{\gamma
1}(\frac{We^{\eta}}{2})$ and $d_{2}=z_{2}\theta_{\gamma
2}(\frac{We^{-\eta}}{2})$ where $\theta_{\gamma}$ is scattering
angle of backscattered photons with respect to the direction of the
incoming electron varies with photon energy
as~\cite{Borden92,telnov90,telnov95}:
\begin{equation}
\theta_{\gamma}\left(\omega\right)\approx\frac{m_{e}}{E_{b}}\sqrt{\frac{E_{b}x}{\omega}-x+1}
\end{equation}
The luminosity spectrum of CLIC $\gamma\gamma$ collider can be seen
in Fig~\ref{fig:lumi.eps}.
\section {FEL System for $\gamma\gamma$ Collider}
\subsection {CLIC main linac, drive linac and master laser}
\begin{figure}
%\begin{center}
%\rotatebox{90}{
\scalebox{0.99999}{
\includegraphics[width=0.99999\textwidth]{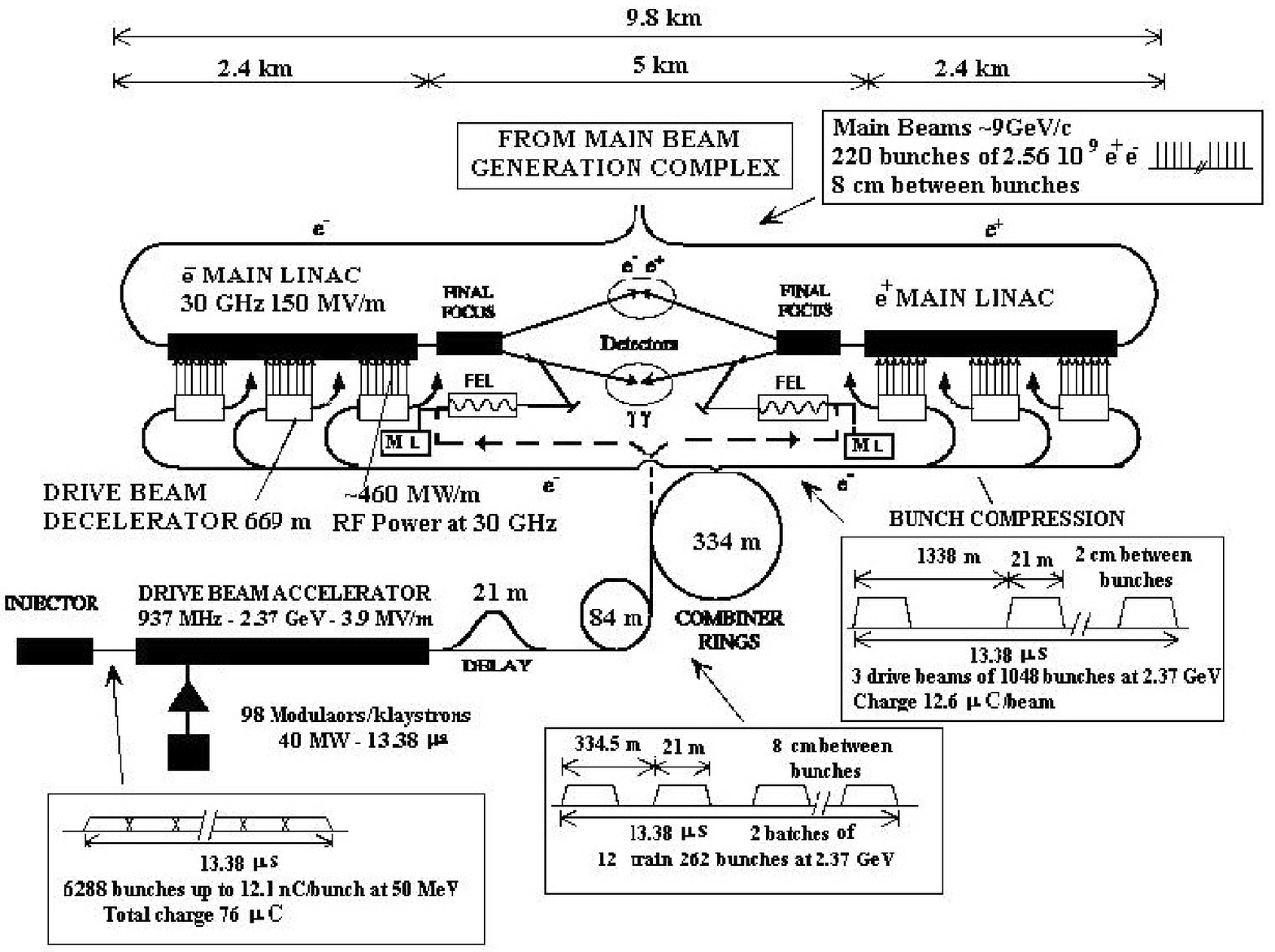}}
\caption{CLIC Layout of $\gamma\gamma$ collider and drive beam
properties at $E_{cm}$=0.5 TeV}\label{fig:clicnma}
%\end{center}
\end{figure}

The time structure of the FEL pulses must follow the time structure
of the electron (positron) bunches of the CLIC main linac at the
conversion point. Using backscattered FEL, synchronization of
$\gamma$ beam and main linac electron (positron) beam can be solved.
The CLIC drive beam complex consist of 2 combiner rings and a delay
loop. Each combiner ring compress to drive beam 4 times and the
delay loop another factor of 2. In order to get the same time
structure in the FEL and the main beam it is necessary to use
additional drive beam bunches after the $1^{st}$ combiner ring.
Drive beam bunch structure and it's complex can be seen in
Fig~\ref{fig:clicnma}. In this case both drive beam and main beam
have the same bunch separation (0.267 ns). The number of main beam
bunches per pulse is $220$ and the number of drive beam bunches per
train after $1^{st}$ combiner ring is 262. When we take 2 more drive
beam pulse from the drive linac gun, we would have 524 more bunches
after the $1^{st}$ combiner ring to be used wigglers for both
electron main linac and positron main linac. Before the wiggler it
should be dumped 42 e-bunches for each drive linac section so the
bunch number of taken drive beam after $1^{st}$ combiner ring would
be decreased to 220. The example of scheme under consideration and
the overall CLIC layout with updated beam parameters can be seen in
Fig~\ref{fig:clicnma}. The power of master laser must be higher than
FEL amplifier noise at entrance of the
wiggler~\cite{saldin2000}. Furthermore the master laser has to be synchronize with drive linac electron bunches after $1^{st}$ combiner ring.\\
{\em Luminosity increment method}: CLIC at $E_{cm}$=3 TeV
containment 21 drive beam decelerator units for each main linac
section, but the $E_{cm}$=0.5 TeV option it has only 3 decelerator
units. In this case the drive beam pulse length reduced by a factor
of 7. To increase the luminosity, $7\times220$ main beam bunches can
be accelerated by using full drive beam pulse length (93.7 $\mu s$).
Consequently the repetition frequency of main linac increases by
factor 7. CLIC main linac beam parameters are given in
Table~\ref{table1}, where proposed modification of repetition rate
and Luminosity are given in parenthesis. In this case we need a
kicker which is faster than $E_{cm}$=3 TeV option after the second
combiner ring. The drive beam pulse structure can be seen in
Fig~\ref{fig:clicnma}. Master laser and electron interaction in the
wiggler will be explained in next subsection. The required
parameters for the master laser are given in
Table~\ref{ml&amplifier} and related drive beam parameters are given
in Table~\ref{dbparam}.
\subsection {FEL amplifier}
\begin{figure}
\scalebox{0.999}{
\includegraphics[width=0.95\textwidth]{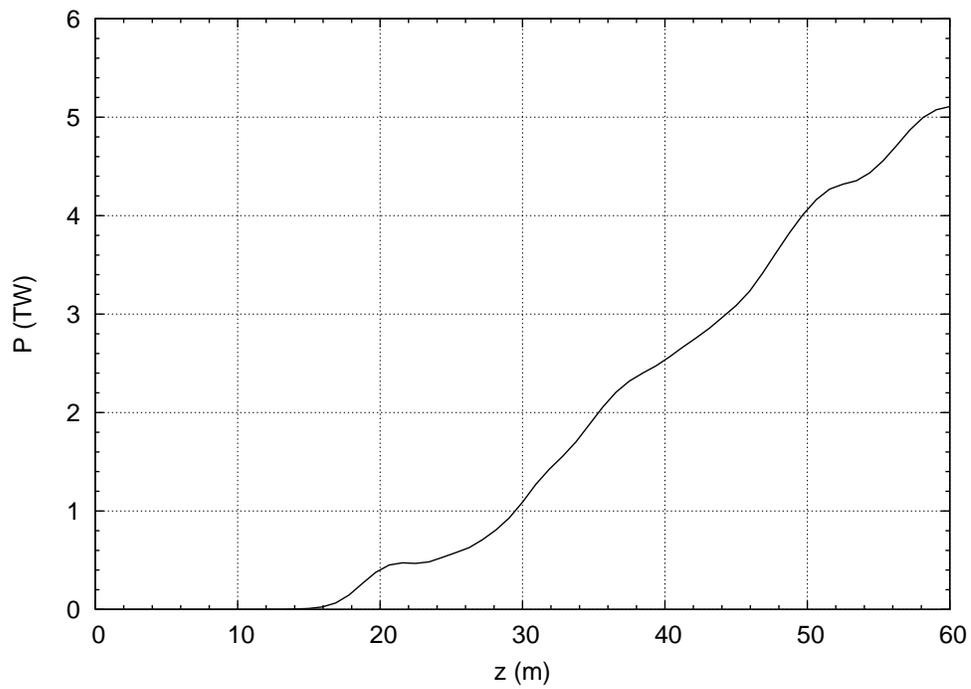}}
\caption{FEL Power as a function of wiggler
length}\label{fig:clic-250.eps}
\end{figure}

\begin{figure}
\scalebox{0.999}{
\includegraphics[width=0.95\textwidth]{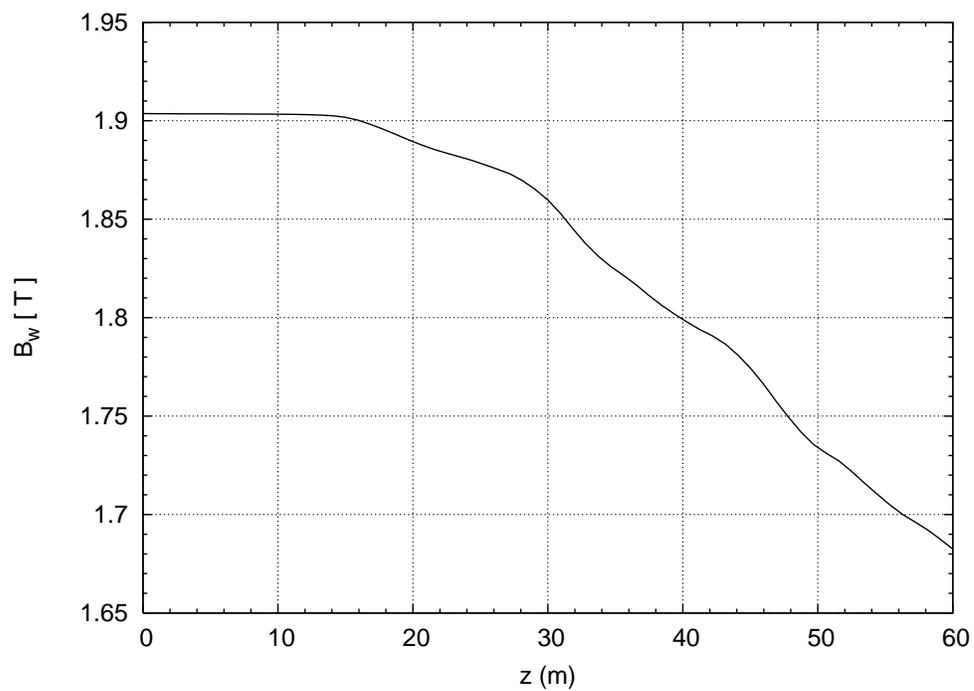}}
\caption{Magnetic field on the wiggler axis vs wiggler
length}\label{fig:clic-250bw.eps}
\end{figure}

Proposed scheme is a high-gain single pass FEL amplifier. The
interaction between electrons and master laser leads to an
exponential growth of the FEL while the electron beam expenses its
kinetic energy. First of all the helical wiggler must satisfy
resonant condition~\cite{saldin95a}:
\begin {equation}
\lambda=\frac{\lambda_{w}}{2\gamma^2}(1+K^{2})
\end {equation}
where $K$=$0.934 B_{w}[T]\lambda_{w}[cm]$ is dimensionless wiggler
parameter, $\lambda_{w}$ is the wiggler period, $B_{w}$ peak
magnetic field inside wiggler and $\gamma$ the drive beam
relativistic factor. The choose of the optimum $\lambda_{w}$,
$B_{w}$ couple can be made by minimizing the gain length $l_{g}$
(e-folding length for the radiation growth). The fundamental FEL
parameter, $\rho=\lambda_{w}/4\pi l_{g}$ corresponds roughly to the
maximum efficiency that can be obtained in a non-tapered FEL; when
this level of efficiency is reached, the loss in kinetic energy of
the electron beam is no longer satisfied, and the output power
saturates. Optimum value of both untapered efficiency and wiggler
length is obtained by minimizing $\lambda_{w}$ and maximizing
$B_{w}$. After saturation, further extraction of energy from the
electron beam is possible by varying the $K$ parameter by changing
wiggler peak field, wiggler period or both. While changing K and the
electrons losses their energy it is possible to keep the electrons
and FEL in resonant. In this case the radiation growth is anyway no
more exponential. Different type of tapering techniques are
possible. In the following we will refer to self-consistent tapering
obtained by varying quadratically the peak wiggler field while
keeping constant the wiggler period. This self-consistent tapering
should be more efficient than the constant $K$. In order to provide
high efficiency it is necessary to use tapered wiggler. FEL has
always tunable and capable to generate powerful coherent radiation
which always has minimal difraction (dispersion). To obtain a
reasonable luminosity of the $\gamma\gamma$ collider at CLIC, the
peak power in the radiation pulse for $p=0.65$ at the FEL amplifier
exit should be $~$0.5 TW. In the case of axisymmetric electron beam
the eigenvalue equation of the $TEM_{mn}$ mode is of the
form~\cite{saldin95a}:
\begin{equation}\label{eigenvalue}
\mu J_{n+1}(\mu)K_{n}(g)=g J_{n}(\mu)K_{n+1}(g)
\end{equation}
where $J_{n}$ is the bessel function of the firs kind of order n,
$K_{n}$ is modified Bessel function, n the azimuthal index of the
mode, $g$ and $\mu$ dimensionless parameter as given
$g$=-$2iB\hat{\Lambda}$ and
$\mu$=$\frac{-2i\hat{D}}{1-i\hat{\Lambda}_{p}^{2}\hat{D}-g^{2}}$,
where $\hat{\Lambda}=\Lambda/\Gamma$ is reduced eigenvalue. For
Gaussian energy spread, the function  $\hat{D}$ and reduced detuning
$\hat{C}$ are defined with formulae
\begin{eqnarray}
\hat{D}=i\int_{0}^{\infty}\xi exp[-\hat{\Lambda}_{T}^{2}\xi^{2}-(\Lambda+i\hat{C})\xi]d\xi\\
\hat{C}=\frac{C}{\Gamma}=(\frac{2\pi}{\lambda_{w}}-\frac{w(1+K^{2})}{2\gamma^{2}c})/{\Gamma}
\end{eqnarray}
where $\omega=2\pi c/\lambda$ is frequency of radiation field.
During the amplification physical effects are connected with the
corresponding dimensionless parameters defining the power of the
effects. These parameters are the diffraction parameter B, space
charge parameter $\Lambda_{p}$, the energy spread parameter
$\hat{\Lambda}_{T}$, and the efficiency parameter $\rho$ as given
respectively below~\cite{saldin2001}:
\begin{eqnarray}
B=\frac{2\Gamma r_{b}^{2} \omega}{c}\\
\hat{\Lambda}^{2}_{p}=\frac{\Lambda^{2}_{p}}{\Gamma^{2}}=\frac{4c^{2}}{[\theta_{w} r_{b}\omega]^{2}}\\
\hat{\Lambda}^{2}_{T}=(\frac{\sigma_{E}^{2}}{E_{0}^{2}}+\frac{\gamma^{4}\sigma_{\theta}^{4}}{(1+K^{2})^{2}})/\rho^{2}\\
\rho=\frac{c\gamma^{2}\Gamma}{\omega(1+K^{2})}=[\frac{I}{I_{A}\gamma}\frac{K^{2}}{1+K^{2}}]^{1/2}
\end{eqnarray}
where $\omega$ frequency of radiation field, $I_{A}$ is Alfven
current which is $\approx$ 17 kA and $\theta_{w}=K/\gamma$ electron
rotation angle. Amplification occur at real part of eigenvalue of
the eigenvalue equation. Energy spread in the electron beam is
assumed to be Gaussian with the rms deviation $\sigma_{E}$. RMS
angle spread given by
$\sigma_{\theta}=\sqrt{\epsilon_{n}/\beta\gamma}$. The gain
parameter $\Gamma$ defines the scale of the field gain and it is
defined as~\cite{saldin2001}:
\begin{equation}\label{}
\Gamma=[\frac{I\omega^{2}{\theta_{w}}^{2}(1+K^{2})^{2}}{I_{A}c^{2}\gamma^{5}}]^{1/2}
\end{equation}
The required laser parameter are given in table~\ref{table1}.
Possible effects during the amplification are:\\
 1){\em Electron beam energy spread}:
If electron beam has an energy spread, the FEL gain can be
depressed, essentially because not all the electrons are exactly
resonant with radiation. If the energy spread is substantially
smaller than the fundamental FEL parameter $\rho$, its effect
negligible.  By detuning initially the beam with respect to the
exact resonance, one can minimize the effect of
the energy spread on the FEL gain~\cite{saldin93,corsini94}.\\
 2) {\em Emittance effect}: There are different ways in which the
electron beam emittance can effect the FEL gain. First of all
transverse motion of the electron in a finite emittance beam induces
a spread in the longitudinal velocities that affects the coupling,
exactly as does the energy spread. FEL gain loss as a function of
energy spread, taking into emittance effects and for optimum
detuning. $(0.005\leq(\gamma-\gamma_{r})/\gamma_{r}\leq 0.02)$
Values of energy spread as big as 1$\%$ can be still tolerated
~\cite{corsini94}. The FEL efficiency parameter $\rho$ and wiggler
length given in table~\ref{ml&amplifier} are all calculated
consistently
taking into account the values of emittance and energy spread reported therein.\\
3) {\em Diffraction losses}: The electron beam emittance determines
the electron beam size in the wiggler, once fixed $\lambda_{w}$ and
$B_{w}$. The wiggler provides a focusing force in both planes, with
a betatron wavelength
$\lambda_{\beta}=\sqrt{2}\lambda_{w}/\theta_{w}$. A matched electron
beam has therefore a constant radius along the wiggler:
$r_{b}=\sqrt{{\epsilon_{r}^{n}\lambda_{\beta}}/(2\pi)}$. In order to
obtain the maximum gain, the transverse section of the electron beam
and the input radiation pulse should exactly overlap, but
diffraction limits the distance over which the light beam have the
same transverse section as the electron beam. This distance is
roughly given by the Rayleigh range $Z_{R}$ of a Gaussian light beam
with a waist of the same section of the electron beam : $Z_{R}=\pi
r_{b}^{2}/\lambda$. Fortunately it is not necessary for the Rayleigh
range be equal or greater than wiggler length in order to preserve
the gain. If $Z_{R}$ is lower or of the order of gain length, the
loses in the effective radiation power seen by electron beam are
compensated by gain (gain guiding effect). Furthermore, the
refractive guiding effect, due to light phase shift in an FEL
high-gain amplifier is dominant when the gain length is shorter than
Rayleigh range, and compensate diffraction losses by confining the
light in the proximity of the electron beam that act essentially
like an optical fiber. The ratio between Rayleigh range and gain
length should be so close to one to avoid gain deterioration by
diffraction losses. Anyway, optical guiding (gain+diffraction
guiding) effects depends on the electron beam radial density
distribution and on the initial conditions for the light
beam~\cite{corsini94}. An exact evaluation of these phenomena done by GINGER 3D simulation code~\cite{ginger}.\\
4) {\em Slippage effect}: The difference in longitudinal velocity
between light pulse and electron beam pulse (slippage effect) can
also effect the FEL instability in ways that may be undesirable,
including lengthening of the amplified light pulse. Other means that
group velocity of radiation in the electron beam is less than the
velocity of light. For a FEL operating in the optical region, the
difference in path length between the light an electron pulses at
the end of wiggler is simply given by the relation $\bigtriangleup
l=\lambda N_{w}$, where $N_{w}$ is number of wiggler period. The
main condition $\bigtriangleup l<<l_{b}$ is satisfy, where $l_{b}$
is drive beam electron pulse length. As mentioned before, to obtain
the needed efficiency the use of a tapered wiggler section
unavoidable. An
evaluation have been made based on 3D calculations.\\
Parameters of the FEL amplifier with included all effects to obtain
0.51 TeV power with tapered wiggler are presented in
table~\ref{ml&amplifier}. In Fig~\ref{fig:clic-250bw.eps} shows
wiggler peak field variation along the wiggler length, it plotted
for self-consistent tapering wiggler. The firs part of wiggler up to
18 m (slightly before saturation) has a constant magnetic field then
the peak field start to decrease. In Fig~\ref{fig:clic-250.eps} FEL
pulse power is plotted as a function of wiggler length, it can be
seen from this figure 60 m wiggler is enough to get required 0.51 TW
peak power of FEL. The initial exponential growth of radiation
energy is evident until tapering is introduced. In this case the
final value of the magnetic field is $~$1.67 T. Higher input power
help to decrease wiggler length.\\
\section {Conclusion}
Present paper indicate that CLIC project can solve laser
requirements for $\gamma\gamma$ collider itself. For $x$=4.8
required laser power is $~$0.51 TW and it can be obtained using 60 m
wiggler. Required FEL can be obtain one extra drive beam pulse from
the drive gun after last bunch of CLIC drive beam is deflected into
a helical wiggler after $1^{st}$ combiner ring, which does not loose
any energy at the end of drive linac. A master laser pulse ($\lambda
\sim 1\mu m$) from a solid state laser, synchronized with the drive
beam, is also injected in the helical wiggler. Here the
amplification of the radiation pulse by FEL interaction occurs. The
amplified light (FEL) is directed with the help of proper mirros to
the conversion point, where it interact with the incoming main
electron beam. In this point high-energy gammas are obtained by
Compton backscattering. The high energy gammas interact with the the
gammas obtained in the same way in the other half of the linear
collider. Circular polarize gamma can be obtain using by helical
wiggler. Even including some effect on obtaining required peak power
FEL, it is possible to construct $\gamma\gamma$ or $\gamma e$
collider based on CLIC drive beam FEL. Reachable luminosity of
$\gamma\gamma$ collision is 1.44 $10^{34}$ $cm^{-2}s^{-1}$ and in
the case of using increment method it is value 1.0 $10^{35}$
$cm^{-2}s^{-1}$.
% \appendix
% \section{}
% \label{}
\section {Acknowlengment}
The author would like to thank Dr. Roberto Corsini, Dr. Frank
Zimmermann, Dr. Daniel Schulte, Prof. Dr. Saleh Sultansoy, Prof. Dr.
A Kenan Ciftci, CLIC working group and Assoc. Prof. Dr. Gokhan Unel
for useful discussions.\\
This work supported in part by Turkish Atomic Energy Authority.

\newpage
\section*{Tables}
\begin{table}
    \caption{CLIC Main Linac Beam parameters}\label{table1}
    \begin{tabular}{lcc}\hline
    $E_{b}$ (GeV)                                      &250\\
    No.particle per bunch($10^{9}$)                    &2.56\\
    No. of bunch                                       &220\\
    Repetition frequency $f_{rep}$ (Hz)                &150(1050)\\
    $\beta_{x}$/$\beta_{y}$(mm)                        &2/0.02 \\
    Normalized Emittance ($\mu$m) $\gamma\epsilon_{x}$/$\gamma\epsilon_{y}$ &660/10\\
    $\sigma_{z}$ ($\mu m$)                             &31\\
    Total Luminosity $L_{\gamma\gamma}$                &1.44 $10^{34}$ (1.0 $10^{35}$)\\\hline
    \end{tabular}
\end{table}
\begin{table}
    \caption{Drive beam parameters}\label{dbparam}
    \begin{tabular}{lcc}\hline
    $E_{db}$ (GeV)           &2.37 \\
    Peak current (kA)        &3.62 \\
    Bunch length (mm)        &0.4 \\
    Bunch sep (ns)           &0.267\\
    Bunch charge (nC)         &12.1\\
    $\beta_{x}$ $/$ $\beta_{y}$ (mm)  &2/0.2\\
    No. of bunch/pulse after $1^{st}$ CR   &262 (1834)\\
    Repetition frequency $f_{rep}$ (Hz)    &150\\
    Normalized emittance, rms ($\mu$m rad) &150\\
    Bunch separation (ns)                  &0.267\\
    No. bunches $/$ train      &262$/$24 (262$/$168)\\
    Energy spread $\sigma_{E}/E_{db}$      &0.1$\%$\\
    Pulse duration ($\mu$s)       &13.22 (93.7)\\\hline
       \end{tabular}
\end{table}
\begin{table}
    \caption{Master laser, FEL and wiggler parameters}\label{ml&amplifier}
    \begin{tabular}{lcc}\hline
   {\em Master laser}              &\\
   Power (MW)                      &1\\
   wavelength($\mu$m)           &1.06\\
   {\em FEL amplifier}\\
   wiggler Type                 & helical\\
   Period(cm)                      &10 \\
   Length of wiggler (m)           &60\\
   FEL parameter $\rho$            &2.2 $10^{-2}$\\
   Entrance magnetic field (T)     &1.908\\
   Rayleigh length (cm)            &1.7\\
   FEL Power (TW)                  &0.51\\\hline
\end{tabular}
\end{table}
\listoffigures
\end{document}